\documentclass[runningheads]{llncs}
\usepackage{booktabs}
\usepackage[table,xcdraw]{xcolor}
\newcolumntype{Y}{>{\centering\arraybackslash}X}

\usepackage{pifont}
\usepackage{tabularx}
\usepackage{colortbl}
\usepackage[pagebackref,breaklinks,citecolor=citecolor,colorlinks]{hyperref}
\definecolor{citecolor}{RGB}{65,105,225}
\usepackage{multirow}
\usepackage{microtype}
\usepackage{tabularx}
\usepackage[T1]{fontenc}
\usepackage{authblk}
\usepackage{xspace}
\usepackage{graphicx}
\begin{document}
%
\def\x{{\mathbf x}}
\def\L{{\cal L}}
\makeatletter
\DeclareRobustCommand\onedot{\futurelet\@let@token\@onedot}
\def\@onedot{\ifx\@let@token.\else.\null\fi\xspace}
\def\eg{\emph{e.g}\onedot} \def\Eg{\emph{E.g}\onedot}
\def\ie{\emph{i.e}\onedot} \def\Ie{\emph{I.e}\onedot}
\def\cf{\emph{c.f}\onedot} \def\Cf{\emph{C.f}\onedot}
\def\etc{\emph{etc}\onedot} \def\vs{\emph{vs}\onedot}
\def\wrt{w.r.t\onedot} \def\dof{d.o.f\onedot}
\def\etal{\emph{et al}\onedot}
\makeatother

\title{PolyGlotFake: A Novel Multilingual and Multimodal DeepFake Dataset} 
%
%
\author{Yang Hou\thanks{Corresponding author: hou.yang.549@s.kyushu-u.ac.jp},
Haitao Fu, 
Chuankai Chen, 
Zida Li,  
Haoyu Zhang, \and
Jianjun Zhao
}

\authorrunning{F. Author et al.}
%
\institute{Kyushu University}
\maketitle              
\begin{abstract}
With the rapid advancement of generative AI, multimodal deepfakes, which manipulate both audio and visual modalities, have drawn increasing public concern. 
Currently, deepfake detection has emerged as a crucial strategy in countering these growing threats.
However, as a key factor in training and validating deepfake detectors, most existing deepfake datasets primarily focus on the visual modal, and the few that are multimodal employ outdated techniques, and their audio content is limited to a single language, thereby failing to represent the cutting-edge advancements and globalization trends in current deepfake technologies.
To address this gap, we propose a novel, multilingual, and multimodal deepfake dataset: PolyGlotFake. It includes content in seven languages, created using a variety of cutting-edge and popular Text-to-Speech, voice cloning, and lip-sync technologies.
We conduct comprehensive experiments using state-of-the-art detection methods on PolyGlotFake dataset. These experiments demonstrate the dataset's significant challenges and its practical value in advancing research into multimodal deepfake detection. PolyGlotFake dataset and its associated code are publicly available at: \url{https://github.com/tobuta/PolyGlotFake}

\end{abstract}
\section{Introduction}
In recent years, the emergence of deepfake technology, which leverages advanced deep learning techniques to generate forged content, has captured global attention\cite{juefei2022countering}. A particularly notable significant advancement is the development of multimodal deepfakes\cite{multimodalsurvey}, which manipulate both visual and audio components in videos. This enhancement substantially increases the realism of the forged content, making it increasingly challenging to differentiate from reality.

Recently, the advancement and popularization of cutting-edge technologies such as Text-to-Speech (TTS), voice cloning, and lip-sync have led to the emergence of a new type of multimodal deepfake on the web. Using Platforms like Heygen \cite{heygen2023} and RaskAI \cite{raskai2023}, producers can easily alter the language spoken by characters in videos. creating convincing fake lip-sync videos. This advancement in video tampering technology not only overcomes language barriers but also facilitates the rapid global distribution of deepfake content.\\
\indent The misuse of deepfake technology represents a significant threat to information security. In response, numerous deepfake detection methods have been proposed. These methods \cite{capsule,afchar2018mesonet,efficientVIT,wang2022m2tr} are mainly based on deep learning, and their effectiveness is largely dependent on the quality and diversity of the training data. However, the majority of existing deepfake datasets are unimodal \cite{UADFV,timit,FF++,jiang2020deeperforensics,li2020celeb,FFIW,kwon2021kodf,DF-Platter}, primarily focusing on visual manipulation and often neglecting the audio aspects. Only a few datasets are multimodal \cite{DFDC,khalid2021fakeavceleb}. This scarcity of multimodal deepfake datasets leads to the predominance of visual modality focus in current deepfake detection methods.\\
\indent To the best of our knowledge, DFDC \cite{DFDC} and FakeAVCeleb \cite{khalid2021fakeavceleb} are the only two publicly accessible multimodal deepfake datasets. While these datasets partially meet the demand for multimodal training data, they employ outdated technologies and are predominantly limited to English content. Consequently, they fail to fully represent the global scope and the cutting-edge status of current deepfake technologies, and these limitations could pose generalization challenges in detecting deepfakes. Furthermore, these datasets usually provide only basic attribute labels, like character attributes (\eg, gender), and lack comprehensive labeling of the techniques used.This deficiency makes it difficult to conduct fine-grained technical traceability analysis of the manipulated videos.\\
\indent Considering the global trend and technological advancements of deepfake generation technology, we propose PolyGlotFake, a novel multilingual and multimodal deepfake dataset. Specifically, we collected high-quality videos in seven different languages from publicly available video platforms and translate the content of these video into the six other languages. We employ five advanced voice cloning and TTS technologies to generate audio in the target languages. Then, we employ two cutting-edge lip-sync technologies to produce high-quality, realistic, translated videos. Each video is accompanied by detailed technical and attribute labels, which are crucial for analysis and classification in technical traceability. Furthermore, we conduct a comprehensive evaluation of current state-of-the-art deepfake detection methods on our dataset. Experimental results demonstrate the challenges of PolyGlotFake in deepfake detection tasks and its practical value in advancing multimodal deepfake detection research.\\
\indent Our contributions are summarized as follows:
\renewcommand{\labelitemi}{$\bullet$}
\begin{itemize}
  \item We present a novel multimodal, multilingual deepfake dataset comprising seven languages and created using ten multimodal manipulation methods. Notably, no multilingual deepfake dataset has been proposed previously.
  \item We carefully selected raw videos in seven languages from public platforms and annotated each with fine-grained labels for character features and specific techniques. This deepfake dataset enables more detailed traceability of the technologies used.
  \item We comprehensively evaluated current state-of-the-art deepfake detection methods on PolyGlotFake and conduct comparative experiments with other datasets. These results demonstrate the challenging nature and the value of PolyGlotFake dataset.
\end{itemize}

\section{Background and Motivation}

\begin{table*}[t]
\centering
\caption{\small Quantitative comparison of PolyGlotFake with existing publicly available video deepfake datasets.}
\resizebox{\textwidth}{!}{%
\begin{tabular}{@{}l|c|c|c|c|c|c|c|c|c@{}}
\toprule
\rowcolor[HTML]{EFEFEF} 
\multicolumn{1}{c|}{\cellcolor[HTML]{EFEFEF}DataSet} & Release Data & \begin{tabular}[c]{@{}c@{}}Manipulated \\ Modality\end{tabular} & Mutilingual & Real video & Fake video & Total video & \begin{tabular}[c]{@{}c@{}}Manipulation \\ Methods\end{tabular} & \begin{tabular}[c]{@{}c@{}}Techniques\\ labeling\end{tabular} & \begin{tabular}[c]{@{}c@{}}attribute\\ labeling\end{tabular} \\ \midrule\midrule
UADFV \cite{UADFV} & 2018 & V & No & 49 & 49 & 98 & 1 & No & No \\
TIMI \cite{timit} & 2018 & V & No & 320 & 640 & 960 & 2 & No & No \\
FF++ \cite{FF++} & 2019 & V & No & 1,000 & 4,000 & 5,000 & 4 & No & No \\
DFD \cite{FF++} & 2019 & V & No & 360 & 3,068 & 3,431 & 5 & No & No \\
DFDC \cite{DFDC} & 2020 & \cellcolor[HTML]{FD6864}A/V & No & 23,654 & 104,500 & 128,154 & 8 & No & No \\
DeeperForensics \cite{jiang2020deeperforensics} & 2020 & V & No & 50,000 & 10,000 & 60,000 & 1 & No & No \\
Celeb-DF \cite{li2020celeb} & 2020 & V & No & 590 & 5,639 & 6,229 & 1 & No & No \\
FFIW \cite{FFIW} & 2020 & V & No & 10,000 & 10,000 & 20,000 & 1 & No & No \\
KoDF \cite{kwon2021kodf} & 2021 & V & No & 62,166 & 175,776 & 237,942 & 5 & No & No \\
FakeAVCeleb \cite{khalid2021fakeavceleb} & 2021 & \cellcolor[HTML]{FD6864}A/V & No & 500 & 19,500 & 20,000 & 4 & No & Yes \\
DF-Platter \cite{DF-Platter} & 2023 & V & No & 133,260 & 132,496 & 265,756 & 3 & No & Yes \\ \midrule
PolyGlotFake  & 2023 & \cellcolor[HTML]{FD6864}A/V & Yes & 766 & 14,472 & 15,238 & 10 & Yes & Yes \\ \bottomrule
\end{tabular}%
}
\label{tab:Statistic of fake dataset}
\end{table*}
In this section, we conduct a comprehensive comparison with existing deepfake datasets and detail the limitations of these current datasets. We present a comprehensive list of widely used and publicly available deepfake video datasets for deepfake detection in Table \ref{tab:Statistic of fake dataset}. These datasets reflect the gradual evolution of deepfake video generation techniques.

The early deepfake datasets, such as UADFV \cite{UADFV} and TIMIT \cite{timit}, were created using initial versions of deepfake generation technologies like FakeApp \cite{deepswap2023} and FaceswapGANs \cite{shaoanlu_faceswapgan}. These early datasets are limited in size, contained a small number of low-quality videos, and suffere from significant visual artifacts. Subsequent studies \cite{FF++,li2020celeb} utilized advanced deepfake generation algorithms, targeting creating more diverse and higher-quality deepfake videos with reduced artifacts. Concurrently, several large-scale deepfake datasets \cite{DFDC,jiang2020deeperforensics,FFIW,kwon2021kodf,DF-Platter} have been proposed. However, most of these datasets primarily concentrate on visual modalities, focusing on techniques such as face swapping while neglecting the manipulation of audio modalities. 

Building on previous work, the DFDC \cite{DFDC} dataset emerged as the first multimodal deepfake dataset, incorporating voice cloning in some videos via TTS Skins \cite{polyak2019tts}. However, DFDC's main emphasis is on visual manipulations, and it does not provide clear labeling for audio manipulations, making it difficult to identify which clips have been audio-manipulated. Subsequently, in 2021, FakeAVCeleb \cite{khalid2021fakeavceleb}  was proposed. This dataset includes four types of multimodal forgeries and provides fine-grained labels for each video. While FakeAVCeleb currently stands as the most prominent multimodal deepfake dataset, it faces limitations, notably in the diversity of manipulation techniques and the linguistic variety of the raw videos. It relies solely on SV2TTS \cite{jemine2019real} for audio manipulation, a system considered somewhat outdated, resulting in lower-quality voice synthesis compared to cutting-edge TTS technologies. For lip-sync, it uses an older version of Wav2Lip \cite{wav2lip}, which can produce noticeable artifacts. Another significant limitation is that its real videos are collected from the VoxCeleb2 dataset \cite{chung2018voxceleb2}, which is limited to English, thereby restricting the linguistic diversity available for multilingual deepfakes. These constraints diminish the dataset's variety and realism, impacting the generalizability of detectors trained with it.

As a result, current multimodal datasets still exhibit significant limitations in terms of manipulating technical and linguistic diversity. This research gap highlights the urgent need for more technologically advanced, diverse, and globally representative deepfake datasets.

Furthermore, it is worth noting that many current datasets are often promoted based on their large scale. However, for the specialized task of deepfake detection, an excessively large scale can result in longer training periods. This not only reduces experimental efficiency but may also hinder the ability to quickly iterate and test new detection techniques. Additionally, ensuring the quality and consistency of each sample in a very large dataset can be challenging, which in turn affects the performance and reliability of the model. Therefore, in PolyGlotFake, our emphasis is on creating a high-quality, diverse dataset rather than merely focusing on its scale.

\begin{figure}[t]
    \centering
    \begin{minipage}[b]{0.48\linewidth}
        \includegraphics[width=\linewidth]{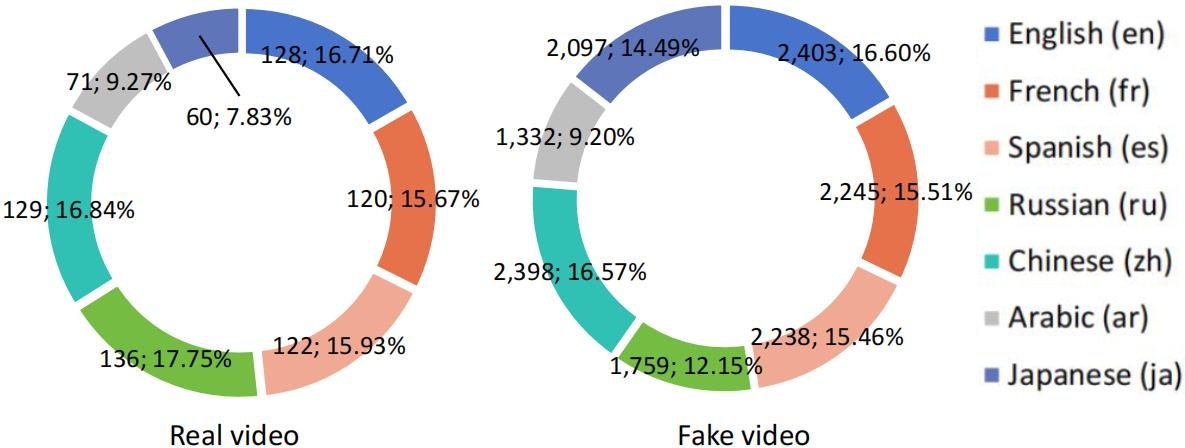}
        \caption{\small Language distribution in real and fake videos.}
        \label{fig:language}
    \end{minipage}
    \hfill
    \begin{minipage}[b]{0.48\linewidth}
        \includegraphics[width=\linewidth]{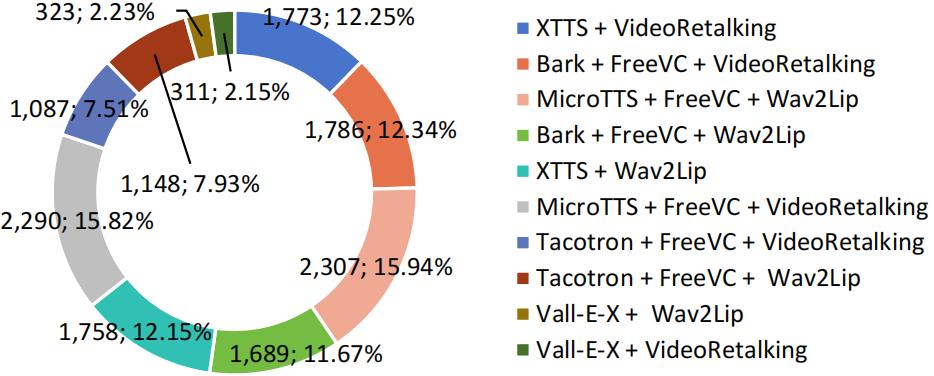}
        \caption{\small Synthesis methods distribution in the PolyGlotFake dataset.}
        \label{fig:tech}
    \end{minipage}
\end{figure}

\begin{table}[t]
\centering
\caption{\small Attribute  distribution by age and sex.}
\begin{tabularx}{\textwidth}{@{}X|Y|Y|Y@{}}  %
\toprule
\multicolumn{2}{c|}{\cellcolor[HTML]{EFEFEF}Characteristics} &
  \cellcolor[HTML]{EFEFEF}Number &
  \cellcolor[HTML]{EFEFEF}Percentage(\%) \\ \midrule\midrule
\multicolumn{1}{c|}{\multirow{4}{*}{Age}} & 0-18   & 2   & 0.26 \\
                                           & 19-35  & 366 & 47.78 \\
                                           & 36-55  & 320 & 41.78 \\
                                           & 56+    & 78  & 10.18 \\ \midrule
\multicolumn{1}{c|}{\multirow{2}{*}{Sex}} & Female & 481 & 62.8  \\
                                           & Male   & 285 & 37.2  \\ \bottomrule
\end{tabularx}
\label{tab:subject}
\end{table}
\section{PolyGlotFake Dataset} 
The PolyGlotFake dataset comprises a total of 15238 videos, including 766 real videos and 14472 fake videos. The average duration of each video is 11.79 seconds, with a resolution of 1280*720.
\subsection{Data Collection}
The high-quality raw (\ie real) videos are collected from YouTube, including content in seven different languages. Figure \ref{fig:language} shows the linguistic distribution in collected raw videos and manipulated videos. To ensure the accuracy of subsequent translations, we manually verify that each sentence in the videos is complete. The selection of languages is based on their global popularity and compatibility with existing popular open-source TTS models. These languages include the six official languages of the United Nations (\ie, English, French, Spanish, Russian, Chinese, Arabic ) and Japanese. We also conducted detailed labels of the collected videos, encompassing information such as their sources, duration, as well as the gender and age of the characters in videos. The attribute distribution by age and sex is shown in Table \ref{tab:subject}. Additionally, we preserved the video's background instead of extracting only facial regions, thereby retaining as much of the original video information as possible.\\
\begin{figure}[t]
    \centering
    \includegraphics[width=\linewidth]{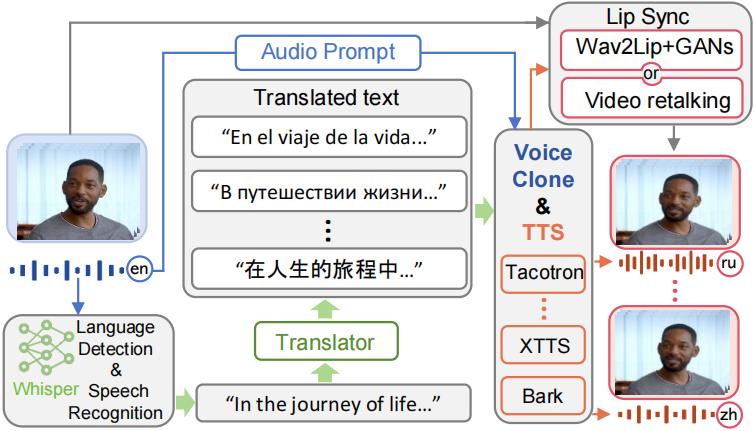}
    \caption{\small Generation Pipeline of PolyGlotFake Dataset. Original videos are separated into video and audio. The audio is transcribed into text using Whisper \cite{openai_whisper} and subsequently translated into multiple languages using a translator. These translated texts are then converted into audio through Text-to-Speech and voice cloning models. Finally, the original video clips are synchronized with the generated audio using a lip-sync model.}
    \label{fig:pipeline}
\end{figure}
\subsection{Synthesized Data}
For the generation of fake videos, we employ cutting-edge and popular visual and audio manipulation methods based on realistic deepfake generation cases found in internet media.

For audio modality manipulation, we use the following five methods.
\begin{itemize}
\item \textbf{XTTS} \cite{coqui_tts_github}: A powerful and popular open-source TTS model built on the Tortoise and developed by Coqui AI. XTTS supports 16 languages and enables cross-lingual voice cloning and multilingual speech generation with only three-second audio prompts. 
\item \textbf{Bark} \cite{bark_suno_ai} + \textbf{FreeVC} \cite{li2023freevc}: Bark is a Transformer-based multilingual TTS model developed by Suno-AI that supports 13 languages and is capable of generating highly realistic, multilingual speech and other audio content such as music. FreeVC is a high-quality, text-free, one-short voice conversion system. Since Bark does not support cross-language voice clones, we use Bark to generate the corresponding speech first and then FreeVC to realize the voice clone according to the audio prompt.
\item \textbf{Vall-E-X} \cite{vall}: An efficient multilingual text-to-speech synthesis and voice cloning model recently proposed by Microsoft. It can efficiently realize high-quality voice cloning with only three seconds of an audio prompt. It currently supports three languages.
\item \textbf{Microsoft TTS} \cite{azure_tts} + \textbf{FreeVC}: Microsoft TTS supports multiple languages and dialects. Given its widespread use on the internet, we design manipulation schemes that combine it with FreeVC.
\item \textbf{Tacotron} \cite{wang2017tacotron} + \textbf{FreeVC}: Tacotron is an advanced TTS synthesis system proposed by Google. It is known for its seq2seq architecture and ability to generate highly natural and fluent speech. Similarly, We combine it with FreeVC.
\end{itemize}

\indent For visual modality manipulation, we employ the following two methods based on the popularity and generation quality:
\begin{itemize}
\item \textbf{Wav2Lip} \cite{wav2lip} + \textbf{GANs}: Wav2Lip is a widely used, highly accurate lip-sync model proposed in 2020. This model can accurately match any speech to the lip movements of a character in a video, often utilized in deepfake for face reenactment tasks. The basic Wav2Lip model alone tends to produce videos of low quality. However, by integrating it with Generative Adversarial Networks (GANs), the video quality can be significantly enhanced. In this study, we employ a fine-tuned Wav2Lip plus GANs model to produce high-quality lip-sync videos.
\item \textbf{VideoRetalking} \cite{cheng2022videoretalking}: VideoRetalking is a audio-driven lip-sync system recently proposed by Cheng \etc. This system generates lip-sync videos by processing audio and video in a series of sequential steps. The generated video frames are finally enhanced and repaired using an identity-aware enhancement network.
\end{itemize}

\indent Additionally, for generated video we label the detailed audio and visual manipulation techniques used, The distribution of the various combinations of techniques is shown in Figure \ref{fig:tech}. For instance, in the pie chart, the gray section represents the percentage of videos that use MicroTTS and FreeVC for voice manipulation, and videoRetalking for lip syncing. There are 2,290 such videos, accounting for 15.82\% of all fake videos.
\begin{table}[t]
\centering
\caption{\small Visual quality assessment and comparison. The first column shows the different Datasets and the second and third columns show the FID and BRISQUE values measured in that Dataset, respectively. lower values of FID and BRISQUE indicate better quality.}
\begin{tabularx}{\textwidth}{@{}Y|Y|Y@{}}
\toprule
\rowcolor[HTML]{EFEFEF} 
DataSet & FID \textdownarrow & BRISQUE \textdownarrow \\ \midrule \midrule
FF++         & 4.12 & 52.17 \\
CelebDF      & 3.72 & 42.23 \\
DFDC         & 5.91 & 74.52 \\
FakeAVCeleb  & 4.32 & 69.31 \\ \midrule
PolyGlotFake & 3.25 & 46.21 \\ \bottomrule
\end{tabularx}%
\label{tab:visual_quality}
\end{table}
\begin{table}[t]
\centering
\caption{\small Audio quality assessment and comparison. The first column shows FakeAVCeleb and the parts of PolyGlotFake that use different sound manipulation techniques. The second column shows the Mos value of the audio in these datasets, where larger indicates higher audio quality.}
\begin{tabularx}{\textwidth}{@{}X|Y@{}}
\toprule
\rowcolor[HTML]{EFEFEF} 
\multicolumn{1}{c|}{DataSet}   & Mos \textuparrow  \\ \midrule\midrule
FakeAVCeleb                   & 3.17 \\ \midrule
PolyGlotFake(XTTS)            & 4.12 \\
PolyGlotFake(MicroTTS+FreeVC) & 4.51 \\
PolyGlotFake(Vall-E-X)        & 3.22 \\
PolyGlotFake(Tacotron+FreeVC) & 4.57 \\
PolyGlotFake(Bark+FreeVC)     & 4.30 \\ 
PolyGlotFake(Overall)         & 4.12 \\ \bottomrule
\end{tabularx}%
\label{tab:audio_quality}
\end{table}

The fake video generation pipeline is shown in Figure \ref{fig:pipeline}. We first extract the audio from the original video and use Whisper \cite{openai_whisper} to convert the speech to text while detecting its language. Then, the text output from Whisper \cite{openai_whisper} is translated into other languages using Microsoft's Translate API. For example, If the output text is in English, the original English text will be translated into Spanish, Russian, Chinese, Japanese, Arabic, and French. We select a suitable TTS model based on the translated text and randomly cut 10 seconds from the original audio as an audio prompt. The selected TTS model converts the text to audio and performs sound cloning based on the audio prompt. Then, the lip-sync model performs face reenactments of the original video based on the TTS output audios, resulting in a series of high-quality manipulated videos in different languages generated using several techniques.
\begin{figure*}[t]
    \centering
    \includegraphics[width=\textwidth]{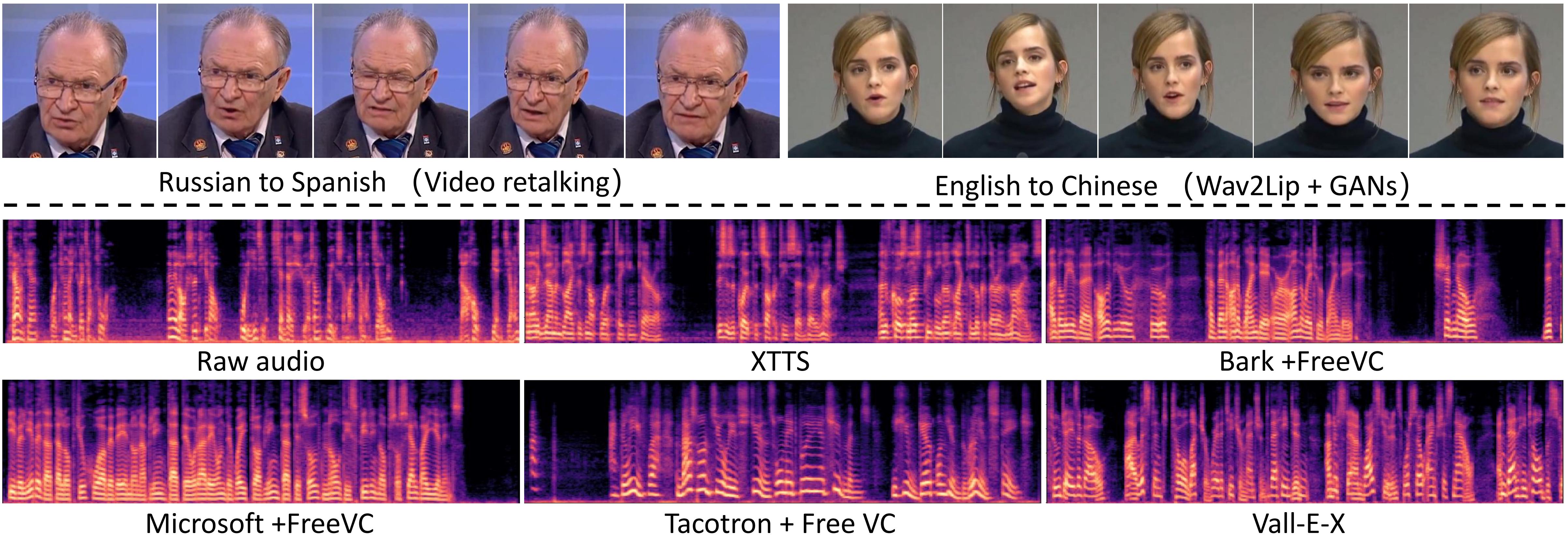}
    \caption{\small Visualization of some video frame samples and Mel spectrograms of audio sample clips in the PolyGlotFake dataset.}
    \label{fig:visual}
\end{figure*}

\subsection{Quality Assessment}
We perform quality assessments for PolyGlotFake dataset in visual and audio modalities. 
For the quality assessment of visual modality, we adopt the Frechet Inception Distance (FID) and the no-reference image assessment method BRISQUE\cite{BRISQUE}. 
We also compar the quality of the PolyGlotFake dataset with several other well-known datasets, including FF++, Celeb-DF, and FakeAVCeleb, and the related results are presented in Table \ref{tab:visual_quality}. For the audio modality quality assessment, we employ the non-invasive audio assessment method NISQA \cite{mittag2021nisqa} to compute the Mean Opinion Score (MOS), and compare the result with FakeAVCeleb. The detailed assessment results for each synthesis method are shown in Table \ref{tab:audio_quality}. 

Based on our quality evaluations, it is clear that the PolyGlotFake dataset exhibits high performance in both visual and audio quality aspects. Additionally, Figure \ref{fig:visual} presents selected video frame samples and Mel spectrograms of audio sample clip from the PolyGlotFake dataset. Both visualization and quantitative quality assessment confirm the superior quality of PolyGlotFake across both visual and audio modalities..

\section{DeepFake Detection Benchmark}
In this section, we comprehensively evaluate several existing state-of-the-art deepfake detectors on the PolyGlotFake dataset and compare the performance of these detectors on different datasets.

\subsection{Selection of Detectors}
Current deepfake detection methods can be broadly categorized into three groups: naive detectors, spatial detectors and frequency detectors. \ding{182} Naive detectors employ CNNs to directly distinguish fake images from real ones. \ding{183} Spatial detectors examine the spatial domain of images in greater detail using specially designed structures to detect features like fusion boundaries and artifacts. \ding{184} Frequency detectors analyze the frequency domain of images to identify forgery features such as high-frequency artifacts. 

To perform the experiments, we employ a total of 13 state-of-the-art deepfake detectors. This set included four naive detectors, namely MesoNet \cite{afchar2018mesonet}, MesoInception \cite{afchar2018mesonet}, Xception \cite{FF++}, and EfficientNet-B4 \cite{tan2019efficientnet};
 five spatial detectors, Capsule \cite{capsule}, FFD \cite{dang2020detection}, CORE \cite{ni2022core}, RECCE \cite{cao2022end}, and DSP-FWA \cite{li2018exposing}; and two frequency detectors, F3Net \cite{qian2020thinking} and SRM \cite{luo2021generalizing}. In addition, for multimodal deepfake detection, we use an ensemble model combining Xception and ResNet, which we call XRes. In this model, Xception is used for visual modality detection, and ResNet is used for audio modality detection. The selection of these detectors was based on the popularity and public availability of their code.
\begin{table}[t]
\centering
\caption{\small Evaluation results and comparisions with other datasets. All detectors were trained on the FakeAVCeleb dataset and tested on FakeAVCeleb, DFDC, and PolyGlotFake. Consequently, the FakeAVCeleb column represents the AUC values obtained from intra-dataset evaluation, while the DFDC and PolyGlotFake columns represent the AUC values from cross-dataset evaluation.}
\resizebox{\textwidth}{!}{%
\footnotesize
\begin{tabular}{c|c|c|cllcllcll}
\hline
\cellcolor[HTML]{EFEFEF} & \cellcolor[HTML]{EFEFEF} & \cellcolor[HTML]{EFEFEF} & \multicolumn{9}{c}{\cellcolor[HTML]{EFEFEF}DataSet}                                    \\ \cline{4-12} 
\multirow{-2}{*}{\cellcolor[HTML]{EFEFEF}Type} &
  \multirow{-2}{*}{\cellcolor[HTML]{EFEFEF}Detector} &
  \multirow{-2}{*}{\cellcolor[HTML]{EFEFEF}Backbone} &
  \multicolumn{3}{c}{\cellcolor[HTML]{EFEFEF}FakeAVCeleb} &
  \multicolumn{3}{c}{\cellcolor[HTML]{EFEFEF}DFDC} &
  \multicolumn{3}{c}{\cellcolor[HTML]{EFEFEF}PolyGlotFake} \\ \hline \hline
Naive                    & MesoNet \cite{afchar2018mesonet}                  & Designed                 & \multicolumn{3}{c|}{0.7332} & \multicolumn{3}{c|}{0.5906} & \multicolumn{3}{c}{0.5672} \\
Naive                    & MesoInception \cite{afchar2018mesonet}           & Designed                 & \multicolumn{3}{c|}{0.7945} & \multicolumn{3}{c|}{0.6344} & \multicolumn{3}{c}{0.5831} \\
Naive                    & Xception \cite{FF++}               & Xception                 & \multicolumn{3}{c|}{0.9169} & \multicolumn{3}{c|}{0.6530} & \multicolumn{3}{c}{0.6052} \\
Naive                    & EfficienNet-B4 \cite{tan2019efficientnet}          & EfficienNet              & \multicolumn{3}{c|}{0.9023} & \multicolumn{3}{c|}{0.6020} & \multicolumn{3}{c}{0.5769} \\ \hline
Spatial                  & Capsule \cite{capsule}                 & Capsule                  & \multicolumn{3}{c|}{0.8663} & \multicolumn{3}{c|}{0.6146} & \multicolumn{3}{c}{0.6068} \\
Spatial                  & FFD \cite{dang2020detection}                     & Xception                 & \multicolumn{3}{c|}{0.9285} & \multicolumn{3}{c|}{0.6583} & \multicolumn{3}{c}{0.5960} \\
Spatial                  & CORE \cite{ni2022core}                    & Xception                 & \multicolumn{3}{c|}{0.9345} & \multicolumn{3}{c|}{0.6625} & \multicolumn{3}{c}{0.6220} \\
Spatial                  & RECCE \cite{cao2022end}                   & Designed                 & \multicolumn{3}{c|}{0.9396} & \multicolumn{3}{c|}{0.6884} & \multicolumn{3}{c}{0.6596} \\
Spatial                  & DSP-FWA \cite{li2018exposing}                 & Xception                 & \multicolumn{3}{c|}{0.9115} & \multicolumn{3}{c|}{0.6929} & \multicolumn{3}{c}{0.6658} \\ \hline
Frequency                & F3Net \cite{qian2020thinking}                  & Xception                 & \multicolumn{3}{c|}{0.9416} & \multicolumn{3}{c|}{0.6452} & \multicolumn{3}{c}{0.6439} \\
Frequency                & SRM \cite{luo2021generalizing}                   & Xception                 & \multicolumn{3}{c|}{0.9043} & \multicolumn{3}{c|}{0.6346} & \multicolumn{3}{c}{0.6143} \\ \hline
Ensemble                 & XRes                     & Designed                 & \multicolumn{3}{c|}{0.9556} & \multicolumn{3}{c|}{0.7042} & \multicolumn{3}{c}{0.6835} \\ \hline
\end{tabular}
}
\label{tab:result}
\end{table}
\subsection{Experimental Setting}
We divide the dataset into training, validation, and testing sets in the ratio of 8:1:1. To ensure the representativeness of each technique combination in the dataset division; we use a stratified sampling method to ensure that the proportion of each combination is consistent across the datasets. For exisiting detection methods, we follow the respective data preprocessing steps. For the ensemble-based model, we randomly clip three seconds from each audio and convert it into a three-channel Mel Frequency Cepstral Coefficient (MFCC) feature as the input for the audio modality and extract ten frames from each video as input for the visual model.

To ensure fairness, we train all detectors on the FakeAVCeleb dataset and evaluate them on both the DFDC and PolyGlotFake datasets. We use the Area Under the Curve (AUC), a commonly used evaluation metric for deepfake detection, as our experimental metric. 

\subsection{Result and Analysis}

Table \ref{tab:result} reports the results of our experiments. The FakeAVCeleb column shows the intra-dataset detection results, which reveal that the spatial detector with a specialized structural design and the frequency detectors outperform the naive detectors. For instance, the detection result of Xception is 0.9169, while CORE, which also utilizes Xception as a backbone, achieves a result of 0.9345.

The DFDC and PolyGlotFake columns present results obtained from cross-dataset detection. Comparing these results with the intra-dataset detection results indicates significant performance degradation for detectors trained on FakeAVCeleb when faced with unseen Deepfake content. Furthermore, the performance of the detectors on the PolyGlotFake dataset is significantly worse than on DFDC. This suggests that PolyGlotFake includes a wider variety of unknown synthesis techniques, making it a more challenging dataset for these detectors.

\section{Conclusion}
In this study, we propose PolyGlotFake, a multilingual, multimodal deepfake dataset that employs cutting-edge multimodal manipulation techniques. Each technique used in this dataset is meticulously annotated to aid in technical traceability analysis.
Furthermore, we comprehensively evaluate various state-of-the-art deepfake detectors on this dataset. The experiment results demonstrate the challenging nature and practical value of our dataset. 
We comprehensively evaluated various state-of-the-art deepfake detectors using this dataset. The experimental results underscore the challenging nature and the practical value of PolyGlotFake, demonstrating its potential to significantly advance the field of multimodal deepfake detection.

In future research, we aim to enhance the linguistic diversity and scale of our dataset. Additionally, in response to recent studies \cite{hou2023evading,jia2022exploring,neekhara2021adversarial} that have shown how adversarial perturbations can help evade detection, we plan to explore methods for implementing such perturbations in practical scenarios. This includes incorporating subtle adversarial tweaks into both the audio and video components of our deepfake content.

\subsubsection{Ethics Statement}
Access to the dataset is restricted to academic institutions and is intended solely for research use. It complies with YouTube's fair use policy through its transformative, non-commercial use, by including only brief excerpts (approximately 20 seconds) from each YouTube video, and ensuring that these excerpts do not adversely affect the copyright owners' ability to earn revenue from their original content. Should any copyright owner feel their rights have been infringed, we are committed to promptly removing the contested material from our dataset.

\subsubsection{Acknowledgements} This work was supported by JST SPRING, Grant Number JPMJSP2136.

%
%
%
\bibliographystyle{splncs04}
\bibliography{mybibliography}

\end{document}